\documentclass[reprint,aps,prl]{revtex4-2}

\usepackage{dcolumn}% Align table columns on decimal point
\usepackage{bm}% bold math
\usepackage{amsmath, amsthm, amssymb, gensymb, calrsfs, wasysym, verbatim, bbm, color, graphics, geometry}
\usepackage{epsfig}
\usepackage{tikz}
\usepackage{array}
\usepackage{enumerate}
\usepackage{mdframed}
\usepackage{graphicx}
\usepackage{dcolumn}
\usepackage{hhline}
\usepackage{lineno}
\usepackage{multirow}
\usepackage{MnSymbol}
\begin{document}
\title{Precision Mass Measurement of Proton-Dripline Halo Candidate $^{22}$Al}

\author{S. E. Campbell}%
\email{campbels@frib.msu.edu}
\author{G. Bollen}%
\author{B. A. Brown}%
\author{A. Dockery}%
\affiliation{%
 Department of Physics and Astronomy, Michigan State University, East Lansing, Michigan 48824, USA
}%
\affiliation{%
 Facility for Rare Isotope Beams, East Lansing, Michigan 48824, USA
}%
\author{K. Fossez}%
\affiliation{%
 Department of Physics, Florida State University, Tallahassee, Florida 32306, USA
}%
\affiliation{%
Physics Division, Argonne National Laboratory, Lemont, Illinois 60439, USA
}%
\author{C. M. Ireland}%
\author{K. Minamisono}%
\author{D. Puentes}%
\affiliation{%
 Department of Physics and Astronomy, Michigan State University, East Lansing, Michigan 48824, USA
}%
\affiliation{%
 Facility for Rare Isotope Beams, East Lansing, Michigan 48824, USA
}%
\author{A. Ortiz-Cortes}%
\affiliation{%
 Facility for Rare Isotope Beams, East Lansing, Michigan 48824, USA
}%
\author{B. J. Rickey}%
\author{R. Ringle}%
\affiliation{%
 Department of Physics and Astronomy, Michigan State University, East Lansing, Michigan 48824, USA
}%
\affiliation{%
 Facility for Rare Isotope Beams, East Lansing, Michigan 48824, USA
}%
\author{S. Schwarz}%
\author{C. S. Sumithrarachchi}%
\author{A. C. C. Villari}%
\affiliation{%
 Facility for Rare Isotope Beams, East Lansing, Michigan 48824, USA
}%
\author{I. T. Yandow}%
\affiliation{%
 Department of Physics and Astronomy, Michigan State University, East Lansing, Michigan 48824, USA
}%
\affiliation{%
 Facility for Rare Isotope Beams, East Lansing, Michigan 48824, USA
}%

\date{December 18, 2023}

\begin{abstract}
We report the first mass measurement of the proton-halo candidate $^{22}$Al performed with the LEBIT facility's 9.4~T Penning trap mass spectrometer at FRIB. This measurement completes the mass information for the lightest remaining proton-dripline nucleus achievable with Penning traps. $^{22}$Al has been the subject of recent interest regarding a possible halo structure from the observation of an exceptionally large isospin asymmetry [Phys. Rev. Lett. \textbf{125} 192503 (2020)]. The measured mass excess value of $\text{ME}=18\;093.6(7)$~keV, corresponding to an exceptionally small proton separation energy of $S_p = 99.2(1.0)$~keV, is compatible with the suggested halo structure. Our result agrees well with predictions from \textit{sd}-shell USD Hamiltonians. While USD Hamiltonians predict deformation in $^{22}$Al ground-state with minimal $1s_{1/2}$ occupation in the proton shell, a particle-plus-rotor model in the continuum suggests that a proton halo could form at large quadrupole deformation. These results emphasize the need for a charge radius measurement to conclusively determine the halo nature.
\end{abstract}

%\keywords{}%Use showkeys class option if keyword display desired

\maketitle

\section{Introduction}
Nuclei at, and near, the boundaries of nuclear existence serve as excellent probes for evaluating nuclear models given their exotic nature \cite{Vretenar-NucStructExotic, Otsuka-ShellStructExotic, Yamaguchi-NucMasses}. Halo nuclei are one such example; seated at or near the nuclear driplines, their mass distribution extends far beyond the compact core, creating their so-called `halo' structure. Efforts to identify, study, and model these nuclei have provided critical evaluations of nuclear forces (see \cite{Tanihata-HaloOverview, Hammer-HaloEFTReview, Pfutzner-ProtonDripRev} for reviews). 

Studies of proton halo structures are often limited, compared to neutron halos, as their formation is suppressed by the confining effect of the Coulomb barrier. Additional complications arise from low production cross-sections for nuclei approaching the proton dripline. Precision measurements of these nuclei are of particular importance as the proton drip line is sufficiently well known compared to that for neutrons, offering a unique probe of nuclear forces. 

$^{22}$Al was first proposed as a proton-halo candidate resulting from an exceptionally small proton separation energy extrapolated from systematic trends in the mass region \cite{Huang-AME2020} and from calculations of isospin-symmetry breaking effects in the sd-shell region \cite{Kaneko-IsospinBreaking}. Recent excitement pointing to a halo structure in the first $1^{+}$ state of $^{22}$Al stems from an observed isospin asymmetry of $\delta_{\beta} = 2.09(96)$ in the Gamow-Teller $\beta^{+}$ transition from the first excited $1^{+}$ state of $^{22}$Si and with its mirror $^{22}$O $\beta^{-}$ transition \cite{Lee-22AlHalo}. This is by far the largest asymmetry reported for low-lying states. Results, however, contain significant uncertainties propagating from its unmeasured mass. The presence of an $s$-wave halo in $^{22}$Al should favor deformation by increasing the proton $1s_{1/2}$---$0d_{5/2}$ quadrupole coupling. In fact, relativistic mean-field calculations \cite{nl3} and data evaluation \cite{moller16} support large quadrupole deformation of about $\beta_2 > 0.2$ in $^{22}$Al. 
% There might be a typo in \cite{nl3}. Ref.~\cite{moller16} is cited as predicting a value of $\beta_2 = 0.271$ for $^{22}$Al, which coincidentally is $-\beta_2$ for $^{42}$Al.

Due to the role of weak binding in the emergence of halo structures, precise knowledge of the binding energy--and consequentially mass--is paramount \cite{Yamaguchi-NucMasses, Lunney-DetNuclearMass}. In quantum mechanics, the halo can be seen as a leakage of the valence nucleon wave function, which itself depends exponentially on the nucleon separation energy. Precise mass information is also required to extract nuclear charge radii from isotope shift measurements \cite{King-IsotopeShift}. Since the current mass value of $^{22}$Al is based purely on extrapolation of systematic trends of mass values of nearby isotopes \cite{Huang-AME2020}, there is a clear need for a first, precise mass measurement of $^{22}$Al. 

\section{Experimental Method and Analysis}
At the Facility for Rare Isotope Beams (FRIB), the proton-rich $^{22}$Al isotope was produced by projectile fragmentation: a $^{36}$Ar primary beam was sent through the FRIB linear accelerator, ultimately impinging upon a $^{12}$C target of about 8~mm thickness. The cocktail beam was then sent to the Advanced Rare Isotope Separator \cite{Hausmann-ARIS} for separation. Two aluminum degraders of thicknesses 3287~$\mu$m and 1015~$\mu$m, followed by a 1003~$\mu$m 2.67~mrad aluminum wedge prepared the beam for stopping in the Advanced Cryogenic Gas Stopper (ACGS) \cite{Lund-ACGS}. Following mass separation with a magnetic dipole, the extracted 30-keV energy beam showed highest beta activity at A=22 and was sent to the Low Energy Beam Ion Trap (LEBIT) facility. There, the continuous beam was delivered to the Cooler-Buncher \cite{Schwarz-CoolerBuncher}, allowing ions to cool for 10~ms. Ion bunches were purified with a coarse time-of-flight gate before injection into the 9.4~T Penning trap mass spectrometer \cite{Ringle-LEBIT}. In the trap, nearby contaminants were cleaned using the Stored Waveform Inverse Fourier Transform (SWIFT) technique \cite{Kwiatkowski-SWIFT}.

LEBIT uses Penning trap mass spectrometry to determine an ion's mass relative to a well-known calibrant through measurements of their cyclotron frequencies, $\nu_c$:  
\begin{equation} \label{eq:mass_from_ratio}
    R = \frac{ \nu_{c,\textrm{calib}} }{ \nu_{c} } = \frac{ q_\textrm{calib} (m - m_e q) }{ q (m_\textrm{calib} - m_e q_\textrm{calib}) }  , 
\end{equation}
where $m$ and $q$ are the ion atomic mass and charge state respectively. Electron binding energies are negligible as they are orders of magnitude smaller than the measurement precision. 

\begin{figure}[htb] 
    \begin{tikzpicture}
        \draw (0, 0) node[inner sep=0] {\includegraphics[width=\columnwidth] {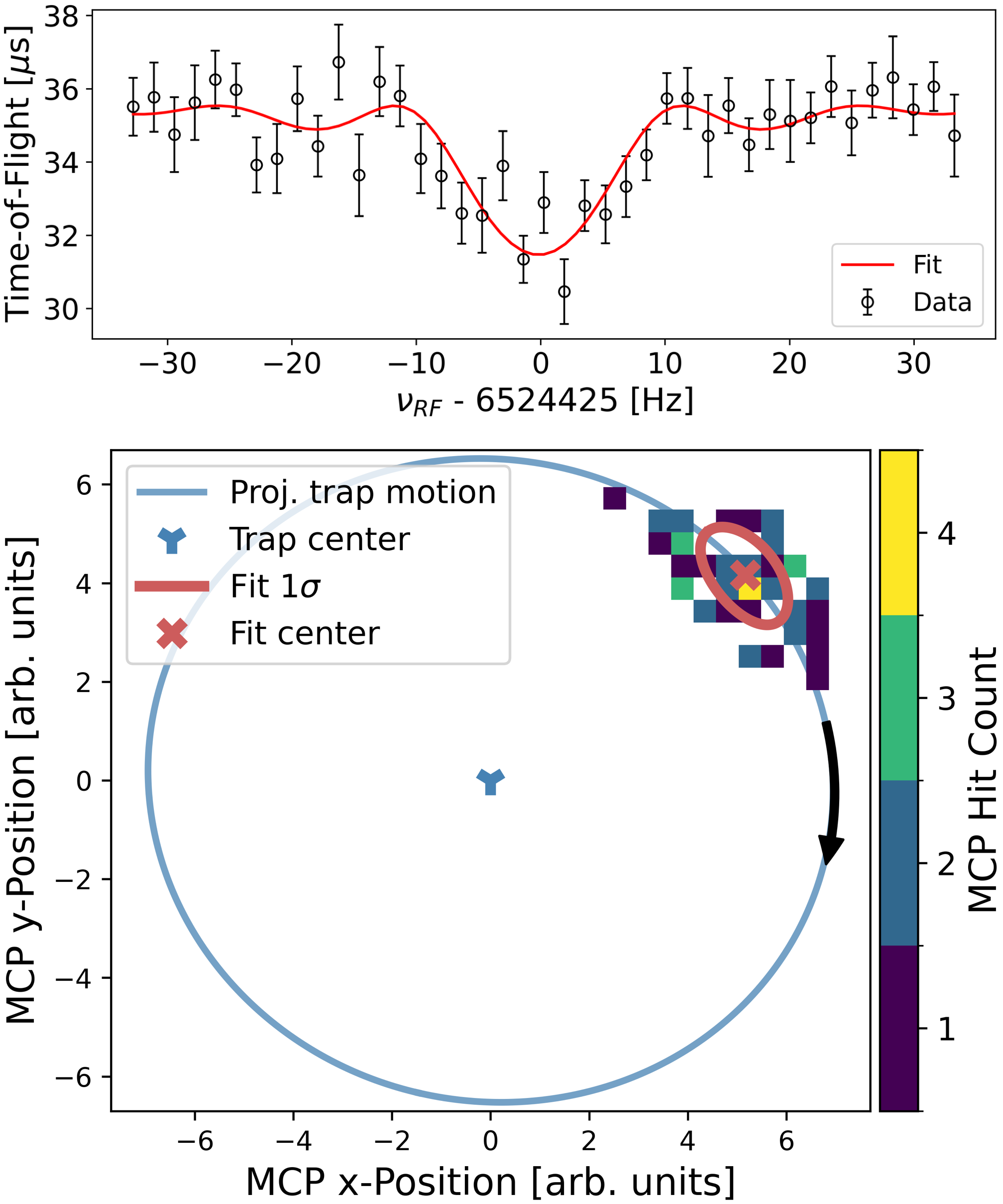}};
        \node[align=left] at (-2.65,2.18) {(a)};
        \node[align=left] at (-2.5,-3.4) {(b)};
    \end{tikzpicture}
    \caption{(a) Observed 75~ms $^{22}$Al$^{+}$ TOF-ICR resonance and fit, containing approximately 350 detected ions. (b) Observed 50~ms PI-ICR beam spot resulting from reduced cyclotron motion, containing approximately 50 detected ions. The spot is fit to a 2D (polar) Gaussian. The black arrow shows the direction of motion. }
    \label{fig:TOF_PI_examples}
\end{figure}

Initial measurements were performed with the well-established Time-of-flight Ion Cyclotron Resonance (TOF-ICR) technique in use at LEBIT \cite{Bollen-TOFICR, Konig-TOFICR, Ringle-LEBIT_PT}. The achievable precision is determined in part by excitation duration, which defines the frequency selectivity and scales linearly with duration. Because of its half-life of $\approx 85$~ms \cite{Wu-Al22HalfLife}, excitation times of 50 and 75~ms were used for $^{22}$Al. A much longer 500~ms excitation was used for $^{23}$Na. Figure~\ref{fig:TOF_PI_examples}(a) shows one TOF-ICR resonance spectrum obtained for $^{22}$Al$^{+}$. 

Additional measurements were performed with the recently implemented Phase Imaging Ion Cyclotron Resonance (PI-ICR) technique \cite{Eliseev-PIICR} at LEBIT. A `two-phase' approach is used, measuring the magnetron ($\nu_{-}$) and reduced cyclotron ($\nu_{+}$) frequencies independently. 
%The excitation pattern used differs slightly from that presented in \cite{Eliseev-PIICR} as LEBIT utilizes Lorentz steerers \cite{Ringle-lorentzSteerer} to precisely center the ions in the trap, removing the need for an excitation to center the ions. 
After evolving for a duration $t_\textrm{acc}$, the final phase $\phi_\textrm{final}$ of the ions relative to a reference location $\phi_\textrm{ref}$ determines the frequency:
\begin{equation}
    \nu_{\pm} = \frac{\left(\phi_{\textrm{final},\pm} - \phi_\textrm{ref}\right) + 2\pi n}{2\pi t_\textrm{acc}} .
\end{equation}
$n$ is the number of full revolutions undergone by the ions during $t_\textrm{acc}$, and is known sufficiently, so long as the frequency uncertainty is less than $1/t_\textrm{acc}$. The ion's cyclotron frequency is found by $\nu_c = \nu_{+} + \nu_{-}$ as described in \cite{Gabrielse-Sideband}. Similar to the restrictions for TOF-ICR, $t_\textrm{acc}$ was set to 50~ms for $^{22}$Al$^{+}$, and 150~ms for $^{23}$Na$^{+}$. Figure~\ref{fig:TOF_PI_examples}(b) shows an example of a final phase spot for $^{22}$Al$^{+}$ with $t_\textrm{acc}=50$~ms.

A number of systematic uncertainties in $\bar{R}$ have been shown to scale linearly with the difference in mass of the calibrant ion and the ion of interest. This includes shifts due to magnetic field inhomogeneities, deviations in the axial alignment of the Penning trap with the magnetic field, and distortions of the electrostatic trapping potential \cite{Bollen-AccHeavyIon}. Mass dependent shifts for TOF-ICR measurements are known to add an uncertainty of $\delta \bar{R} \approx 2 \times 10^{-10} / u$ \cite{Gulyuz-MassDepShift}, which is negligible compared to the statistical uncertainty. Additional contributions, such as those from isobaric contaminants \cite{Bollen-IsomerRes} and space charge effects \cite{Bollen-AccHeavyIon}, were mitigated by in-trap beam purification with SWIFT and by limiting the number of ions in the trap to less than five at a time. Furthermore, studies of non-linear temporal magnetic field instabilities show a contribution to the uncertainty in $\bar{R}$ by no more than $1\times 10^{-9}$ \cite{Ringle-MagStabilitySys} over the course of one hour, which is greater than the duration of any of the measurements performed.

Additional systematic effects were accounted for in the TOF-ICR measurements by performing offline measurements of $^{23}$Na$^{+}$ with $^{22}$Ne$^{+}$ as an isobaric stand-in for $^{22}$Al$^{+}$. Both these ions’ masses are very well known. The ions were produced using an offline plasma ion source operated with high-purity neon gas at LEBIT. Experimental conditions were matched as closely as possible to those with $^{22}$Al, and the total trapping times remained unchanged. All system settings were not modified following the experiment. The systematic uncertainty was determined to be $\delta \bar{R}_\textrm{TOF} = -1(2) \times 10^{-8}$.

With the increased sensitivity of PI-ICR, a number of additional systematic effects are present. Extensive systematic investigations performed at other facilities \cite{Orford-CPTSys, Karthein-PIAnalysis, Nesterenko-JYFLPISys} show that these are dominated by many of the same as listed above for TOF-ICR. PI-ICR specific effects are also largely determined by the electrostatic trapping field harmonicity, magnetic field alignment, and count rate dependent frequency shifts \cite{Chenmarev-TRIGASys}. These are mitigated by using smaller total trapping times, maintaining the same spot location on the MCP, and limiting the MCP count rate to $\approx 2$ ions per shot respectively. A magnetic field misalignment will result in an elliptical projection of the ion motion. This was accounted for by imaging this projection and fitting the result to an ellipse. The acquired fit is shown in Fig.~\ref{fig:TOF_PI_examples}(b). $^{23}$Na and $^{22}$Ne were also used to calculate a systematic uncertainty for PI-ICR, maintaining the accumulation times and angular locations respectively. The systematic uncertainty was determined to be $\delta \bar{R}_\textrm{PI} = 0(2) \times 10^{-8}$. 

\section{Results}
\begin{figure}[htb] 
    \includegraphics[width=\columnwidth] {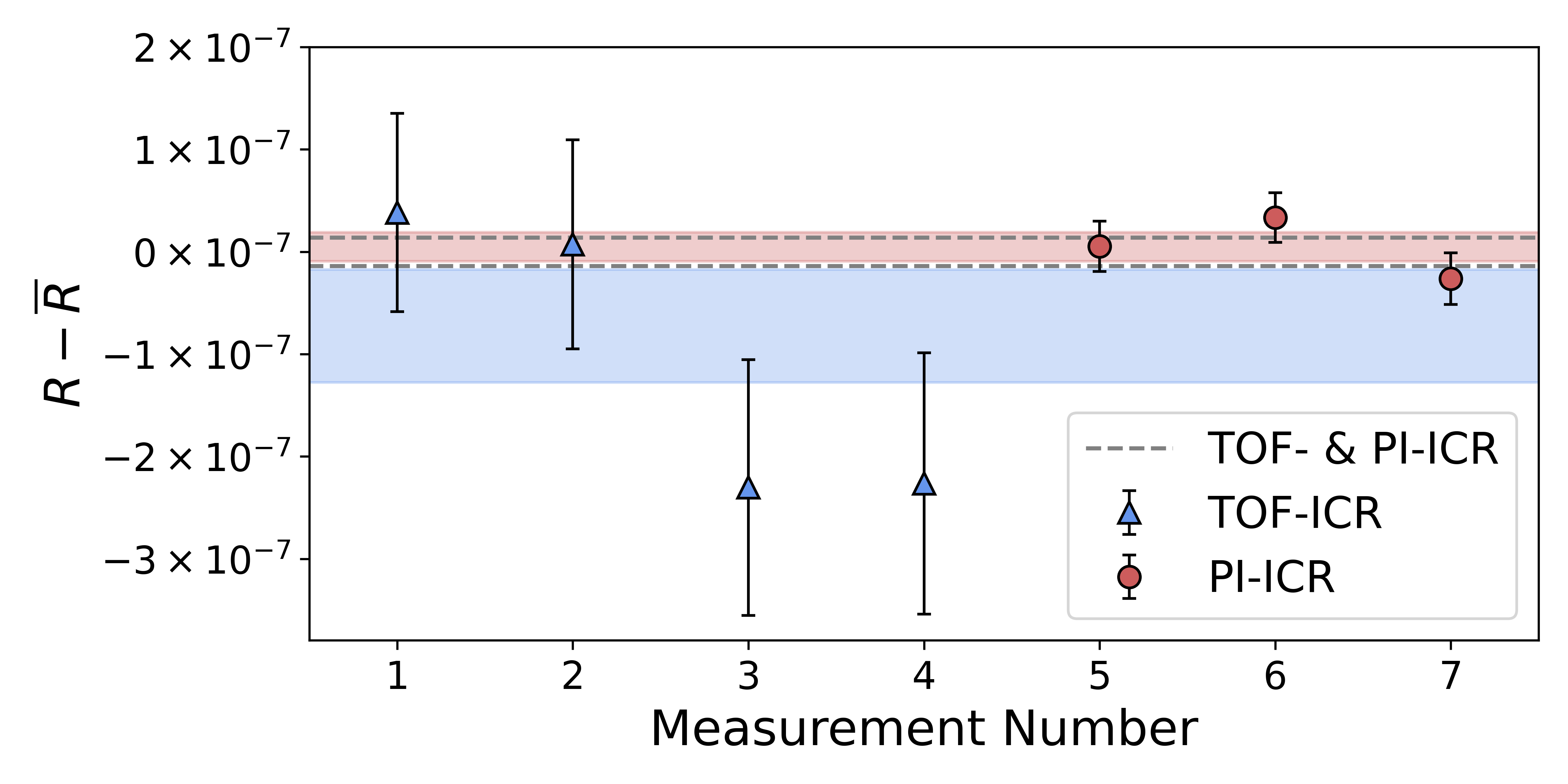}
    \caption{Measured frequency ratios of $^{22}$Al$^{+}$ relative to $^{23}$Na$^{+}$. Blue points and uncertainty region correspond to TOF-ICR measurements, red points and uncertainty region correspond to PI-ICR measurements. Dashed black lines indicate the combined TOF-ICR and PI-ICR result. All uncertainties are shown to $1\sigma$.}
    \label{fig:TOF_PI_measurements}
\end{figure}

A total of four TOF-ICR and three PI-ICR measurements were taken over the course of approximately 16 hours. The TOF-ICR measurements resulted in a weighted average of $\bar{R}_\textrm{TOF} = 1.04406876(6)$, and for the PI-ICR measurements $\bar{R}_\textrm{PI} = 1.044068841(14)$. A summary of these measurements is shown in Fig.~\ref{fig:TOF_PI_measurements}. Combined, these results yield $\bar{R} = 1.044068837(14)$, corresponding to a mass excess of $18\;093.6(7)$~keV and a proton separation energy of 99.2(1.0)~keV.

\section{Discussion}
\subsection{Shell Model with USD Hamiltonians}
\begin{figure}[htb]
    \includegraphics[width=\columnwidth]{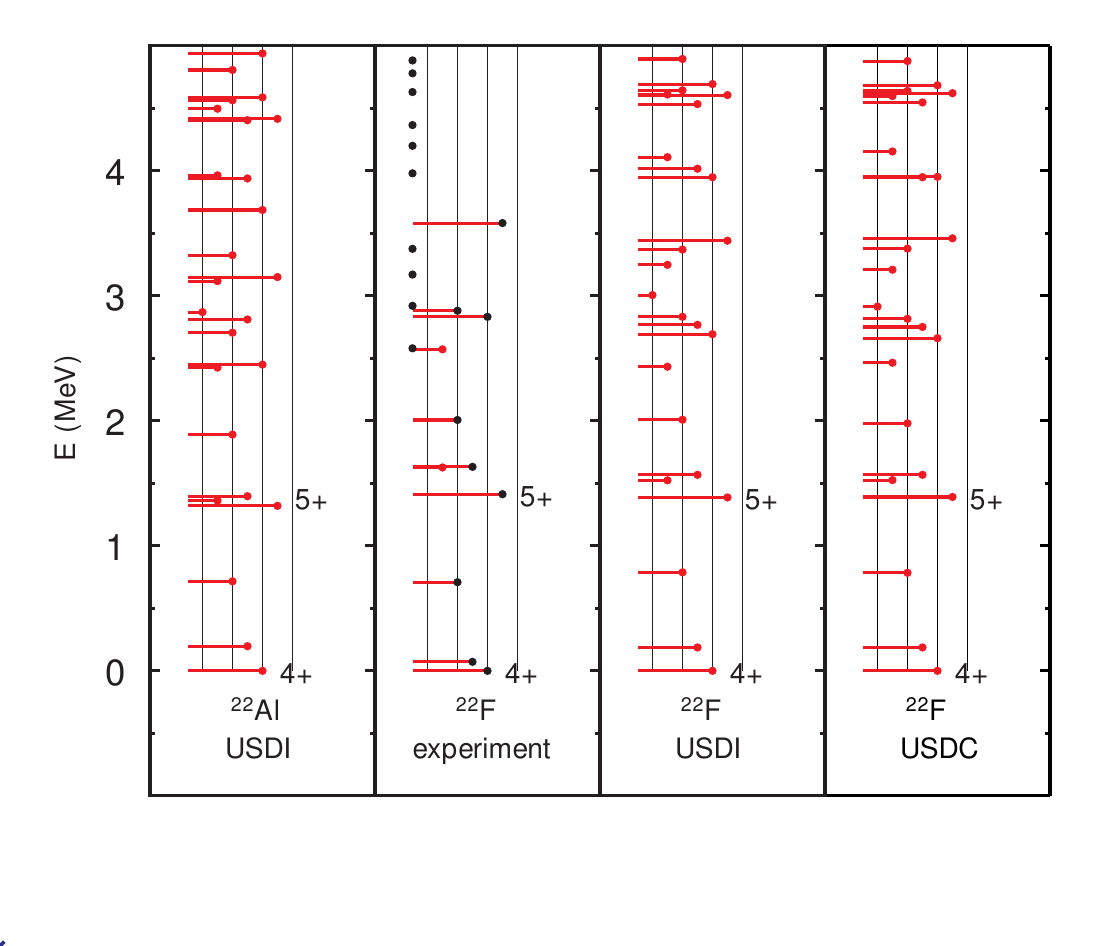}
    \caption{ Experimental levels of $^{22}$F up to 5~MeV compared to the results obtained from the USDC and USDI Hamiltonians. The calculated levels of $^{22}$Al are also shown. The horizontal lines in the figure are proportional to the $J$ value for positive parity states. For the experiment shown on the left-hand side, the levels with unknown  $J$ are shown by black dots. Levels with uncertain $J$ are indicated by the red lines ending with a black dot. }
    \label{fig:22F_spectrum}
\end{figure}

The low-lying structure for the nuclei around $^{22}$Al can be described within the $sd$ $(0d_{5/2},0d_{3/2},1s_{1/2})$ model space. Several ``universal" $sd$-shell (USD) Hamiltonians have been developed over the last 40 years \cite{usd}, \cite{usda}, \cite{usdc}. They are all obtained by starting with a set of two-body matrix elements (TBME) obtained with nucleon-nucleon potentials renormalized to the $sd$ model space.  The singular-valued decomposition (SVD) method is then used to constrain linear-combinations of TBME to obtain lower uncertainties for binding energies and excitation energies. The lastest SVD-type fit from 2020 was based on Hamiltonians that included the Coulomb and isospin-dependent strong interactions treated explicitly \cite{usdc}, and provide absolute binding energies for all $sd$ shell nuclei relative to that of $^{16}$O. This resulted in two Hamiltonians: USDC starting with the Bonn-A \cite{bonna} interaction, and USDI starting with the VS-IMSRG method \cite{imserg1}, \cite{imserg2}. The rms deviation between the experimental and calculated energies for both USDC and USDI was 140~keV. The results for $S_{p}$($^{22}$Al) are 40~keV (USDC) and 30~keV (USDI) which are in reasonable agreement with experiment. The calculated spectrum of the mirror nucleus $^{22}$F is in excellent agreement with experiment (see Fig.~\ref{fig:22F_spectrum}).

As discussed in Sec.~V of \cite{usdc}, a special consideration for proton-rich nuclei is the Thomas-Ehrman shift (TES). For example, the 1/2$^{+}$ excited state in $^{17}$F is shifted down by 0.376~MeV compared to its energy in the mirror nucleus $^{17}$O. Single-particle states with  low-$\ell$ values that are loosely bound (in this case $\ell$=0 for the $1s_{1/2}$ which is bound by only 0.105~MeV in $^{17}$F) are lowered in energy relative to those with high-$\ell$ values. The reason is that the low-$\ell$ states are extended in size, and the kinetic energy of the extended wavefunctions decreases. As discussed in \cite{usdc}, the TES depends on the separation energy of the specific state and the related spectroscopic factors. It is a many-body effect that cannot be corrected simply by a modification of single-particle energies in the Hamiltonian. In the $^{22}$Al mass region there are excited states that have a significant $\ell$=0 spectroscopic factor with a TES as high as $-0.75$~MeV (see Fig.~12 of \cite{usdc}).

A key feature for all of the USD Hamiltonians is the increase in the $1s_{1/2}$---$0d_{5/2}$ single-particle energy gap from 0.9~MeV at $N=8$ to 4.0~MeV at $N=14$ (see Fig.~4 in \cite{bab05}). This shift has been attributed to three-body interactions between the valence neutrons and the core nucleons \cite{ot10} and it is contained in VS-IMSRG Hamiltonian for the $sd$ shell \cite{vs}. Taking into account the TES, the proton $1s_{1/2}---0d_{5/2}$ gap changes from 0.5~MeV at $Z$=8 to about 3.5~MeV at $Z=14$. The calculated proton $1s_{1/2}$ occupations for the $4^{+}$ and $3^{+}$ states of $^{22}$Al are 0.29 and 0.38, respectively. 

The $3^{+}$ state in the mirror nucleus $^{22}$F is only 0.072~MeV above the $4^{+}$ ground-state. The TES from the larger $1s_{1/2}$ occupation of the $3^{+}$ state could result in a $3^{+}$ ground-state for $^{22}$Al, as suggested by \cite{Blank-22Al_3pGndState, Czajkowski-22Al_3pGndState}. However, the calculated $\beta^{+}$ decay properties of $^{22}$Al are more consistent with a $4^+$ ground-state assignment. It will be important to measure the spin-parity and the moments of the $^{22}$Al ground-state to better characterize the structure.

\subsection{Particle-plus-Rotor Model}
Based on the small proton separation energy of about 0.1 MeV measured in $^{22}$Al, one can also investigate the possible presence of a ground-state halo structure using the particle-plus-rotor model (PRM) including couplings to continuum states. The motivation for using this model is to investigate the possible interplay between weak binding, which enhances the occupation of $s$-waves, and deformation via the $s_{1/2}$---$d_{5/2}$ quadrupole coupling, which lowers the proton $1s_{1/2}$ shell. Here, one wants to test how the presence of deformation affects the formation of a halo structure.

We assume that only couplings between the angular momentum $j_r$ associated with the rotational motion of $^{21}$Mg and the angular momentum $j$ of the weakly-bound valence proton play a role in the particle-plus-rotor dynamic ($\vec{j}_r + \vec{j} = \vec{j}_\text{PRM}$), while the angular momentum $J_i$ coming from the structure of $^{21}$Mg can be ignored for our purpose. This approximation is justified by the fact that, except for one well-bound neutron in $0d_{5/2}$, all nucleons in $^{21}$Mg are expected to be paired and coupled to $0^{+}$. It follows that in this picture, the low-lying multiplets $(1,2,3,4,5)^{+}$ and $(2,3)^{+}$ of $^{22}$Al are associated with $j_\text{PRM}^\pi = {5/2}^+$ and ${1/2}^+$, respectively, and the states within each multiplet are degenerate.

In the PRM \cite{fossez16_1335}, the core-nucleon interaction is represented by a Woods-Saxon potential with quadrupole deformation controlled by the deformation parameter $\beta_2$, and the wave function is expanded in terms of channel wave functions defined by the quantum numbers $(n,l,j,j_r)$ where $(n,l,j)$ are the usual shell quantum numbers for the valence particle. Each channel wave function is expanded using the Berggren basis to account for continuum couplings, and the coupled-channel equations obtained are solved exactly by direct diagonalization.

The above shell model calculations using the USD family of interactions, predict a fairly limited occupation of the $\pi 1s_{1/2}$ shell in $^{22}$Al and are not indicative of any significant deformation. Consequently, we first assume spherical symmetry for $^{21}$Mg ($\beta_2=0$) and adjust the potential to reproduce the experimental binding energy of $^{22}$Al and use the energy spectra of $^{21}$Mg to fix the $1s_{1/2}$ shell at about 200~keV above the $0d_{5/2}$ shell, respectively, which will be understood as Nilsson orbits for $\beta_2 \neq 0$. These s.p. energies differ from those in magic nuclei to account for the absence of residual interaction in the PRM. We obtain a diffuseness $d=0.67~\text{fm}$, a radius $R_0 = 3.2~\text{fm}$, a depth $V_0 = 54.8~\text{MeV}$, a spin-orbit coupling $V_\text{so} = 3.62~\text{MeV}$, and a WS Coulomb potential radius $R_c = 3.0~\text{fm}$.

\begin{figure}[htb]
	\includegraphics[width=\linewidth]{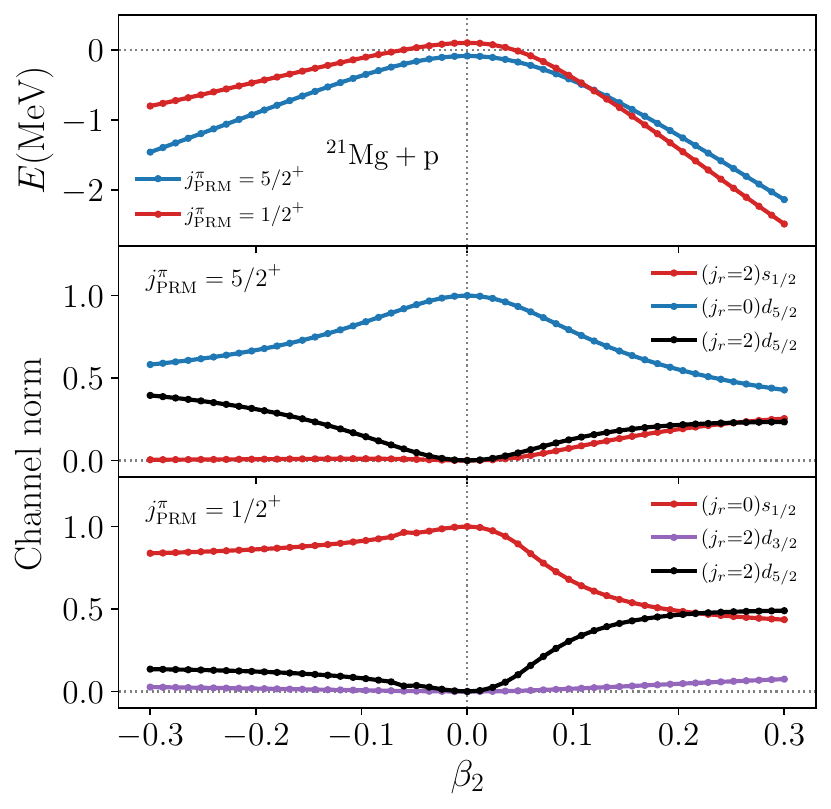}
    \caption{Energy of the $^{21}$Mg+p system (top panel), and norms of channel wave functions for $j_\text{PRM} = {5/2}^{+}$ (middle panel) and ${1/2}^{+}$ (lower panel) as a function of the quadrupole deformation parameter $\beta_2$.}
	\label{fig_PRM}
\end{figure}

The potential is then deformed by varying $\beta_2$, and the ground-state energy of $^{22}$Al together with the norms of channel wave functions are calculated. Results are shown in Fig.~\ref{fig_PRM}.
Both prolate ($\beta_2>0$) and oblate ($\beta_2<0$) deformations are shown to lower the energy, suggesting that a weakly-bound valence proton pushes the system to deform. 

For a small deformation ($|\beta_2|< 0.1$) qualitatively compatible with shell model calculations, the weight of the channel wave function $(j_r=2)1s_{1/2}$ remains relatively low compared to other channels built on the $0d_{5/2}$ shell. In this scenario, the presence of a $s$-wave halo would thus be unlikely. However, for a significant prolate deformation $\beta_2 > 0.1$, the ${1/2}^{+}$ PRM state associated with the $(2,3)^{+}$ multiplet in $^{22}$Al becomes the ground state. This state is dominated by the $s$-wave channel and shown in Fig.~\ref{fig_PRM} (lower panel) and would likely support a halo structure. 
While such large quadrupole deformations are not supported by the shell model,  relativistic mean-field calculations \cite{nl3} and data evaluation \cite{moller16}, however, predict deformations of $\beta_2 = 0.284$ and $0.226$, respectively. 
Using the latter estimate as a guess and readjusting the core-proton potential to match the experimental binding energy by lowering the potential depth does not affect the results significantly.

\section{Conclusions}
In summary, the first precision mass measurement of $^{22}$Al is reported. The exceptionally small proton separation energy observed at 99.2(1.0)~keV agrees with extrapolations by the 2020 Atomic Mass Evaluation as well as predictions made with the \textit{sd}-shell USD Hamiltonians. Such a small nucleon separation energy is one such benchmark allowing for halo formation. The shell model, reproducing the experimental result within uncertainties, suggests a very small $1s_{1/2}$ orbital occupation. In the PRM, this small $1s_{1/2}$ occupation translates in a small deformation parameter and no halo structure. Conversely, the observation of a halo state would indicate strong continuum-induced deformation similar to the suspected situation in $^{29}$F \cite{fossez22_2540}. Both models are constrained by a $4^{+}$ ground state assignment for  $^{22}$Al, though it remains unmeasured \cite{Achouri-22AlGndStateSpin, Wu-Al22HalfLife}. Ultimately, the charge radius and nuclear spin, which can be determined from the hyperfine structure measurements by laser spectroscopy, are needed to conclusively determine the halo nature of the $^{22}$Al ground state and to resolve the confusion surrounding its structure.

\begin{acknowledgments}
    This work was conducted with the support of Michigan State University and the National Science Foundation under Grants No. PHY-1102511, PHY-1126282, PHY-2111185, and PHY-2238752. This material is based upon work supported by the U.S. Department of Energy, Office of Science, Office of Nuclear Physics and used resources of the Facility for Rare Isotope Beams (FRIB) Operations, which is a DOE Office of Science User Facility under Award Number DE-SC0023633, and under the FRIB Theory Alliance Award No. DE-SC0013617. Thank you to Witek Nazarewicz for his fruitful discussions and input towards the presented theoretical work.
\end{acknowledgments}

\bibliography{aluminum22}% Produces the bibliography via BibTeX.

\end{document}